\documentclass[aps,prl,twocolumn,showpacs,superscriptaddress]{revtex4}
\usepackage{graphicx}
\usepackage{epstopdf}
\usepackage{latexsym}
\usepackage{amssymb}
\usepackage{amsmath}
\usepackage{amsfonts}
\usepackage{subfigure}
\usepackage{bm}
\usepackage{multirow}
\usepackage{stmaryrd}

\newcommand{\bra}[1]{\langle #1|}
\newcommand{\ket}[1]{|#1 \rangle}
\newcommand{\db}[1]{\llbracket #1\rrbracket}

\newcommand{\eqnref}[1]{Eq.\,\eqref{#1}}
\newcommand{\figref}[1]{Fig.\,\ref{#1}}

\newcommand{\op}[2]{\vcenter{\hbox{\includegraphics[height=#2]{#1}}}}
\newcommand{\graph}[1]{\vcenter{\hbox{\includegraphics[width=50pt]{#1}}}}

\newcommand{\bpm}{\begin{pmatrix}}
\newcommand{\epm}{\end{pmatrix}}

\begin{document}

\title{Synthetic Topological Degeneracy by Anyon Condensation}
\author{Yi-Zhuang You}
\affiliation{Institute for Advanced Study, Tsinghua University, Beijing, 100084, China}
\author{Chao-Ming Jian}
\affiliation{Department of Physics, Stanford University, Stanford, CA 94305, USA}
\author{Xiao-Gang Wen}
\affiliation{Perimeter Institute for Theoretical Physics, Waterloo, Ontario, N2L 2Y5 Canada}
\affiliation{Department of Physics, Massachusetts Institute of Technology, Cambridge, Massachusetts 02139, USA}
\affiliation{Institute for Advanced Study, Tsinghua University, Beijing, 100084, China}

\date{\today }

\begin{abstract}

Topological degeneracy is the degeneracy of the ground states in a many-body system in the large-system-size limit. Topological degeneracy cannot be lifted by any local perturbation of the Hamiltonian. The topological degeneracies on closed manifolds have been used to discover/define topological order in many-body systems, which contain excitations with fractional statistics. In this paper, we study a new type of topological degeneracy induced by condensing anyons along a line in 2D topological ordered states. Such topological degeneracy can be viewed as carried by each end of the line-defect, which is a generalization of Majorana zero-modes. The topological degeneracy can be used as a quantum memory. The ends of line-defects carry projective non-Abelian statistics, and braiding them allow us to perform fault tolerant quantum computations.

\end{abstract}

\pacs{05.30.Pr, 05.50.+q, 61.72.Lk, 03.67.-a}
\maketitle

\emph{Introduction}--- The topological degeneracy associated to the topological
defects in the topologically ordered states\cite{topo order} has attracted much
research interests
recently\cite{BarkeshliWen,BarkeshliQi,BarkeshliJianQi,Clarke,Lindner,MCheng,Vaezi,Bombin,KitaevKong,YouWen,Ran}.
Examples have been found in $K$-matrix\cite{Kmat} 
$\bigl(\begin{smallmatrix} N & 0\\ 0 & N \end{smallmatrix}\bigr)$
or
$\bigl(\begin{smallmatrix} N & 0\\ 0 & -N \end{smallmatrix}\bigr)$
quantum Hall systems\cite{BarkeshliWen,BarkeshliQi,BarkeshliJianQi,Clarke,Lindner,MCheng,Vaezi},
and in $\mathbb{Z}_N$ gauge theory ($K=\bigl(\begin{smallmatrix} 0 & N\\ N & 0\end{smallmatrix}\bigr)$)\cite{Bombin,KitaevKong,YouWen}, where the topological defects are
rendered into projective non-Abelian anyons\cite{pnAa} due to the interplay
between the space geometry and the topological order. In most of the cases, the
topological defects are realized as lattice dislocations involving non-local
deformation of the lattice structure. The difficulty of manipulating lattice
dislocations hinders the physical realization of fusing or braiding these
projective non-Abelian anyons.  Therefore we are motivated to design flexible
and movable synthetic dislocations. The same motivation also leads to the
proposal of synthetic dislocations in the bilayer FQH systems by zig-zag
gating\cite{gating}. In this work, we show that it is possible to mimic the
lattice dislocations by applying some external field to a line of sites (along
the dislocation branch-cut) without altering the underlying lattice structure.
The external field quenches the degrees of freedom on those sites, as if they
were removed from the lattice effectively. As the external field can be turned
on and off, the synthetic dislocations can be produced and moved around
physically.  On the other hand, the external field (or coupling) also drives
intrinsic anyon condensation\cite{KitaevKong, anyoncond} along the branch-cut
line, which demonstrates the designing principle of synthetic dislocations by
anyon condensation. This makes synthetic dislocations more general than the
lattice dislocations, since we can choose to condense different types of
anyons, which in turn generates different kinds of topological degeneracies,
associated to different synthetic dislocations with different projective
non-Abelian statistics.




\emph{Synthetic Dislocations}--- We start from the $\mathbb{Z}_2$ plaquette
model\cite{Wen plaq} (or the Kitaev toric code model\cite{Kitaev}), whose low
energy effective theory is a $\mathbb{Z}_2$ lattice gauge theory.  It has been
shown\cite{YouWen} that each lattice dislocation in this model is associated to
a Majorana zero mode. On the other hand, the $\mathbb{Z}_2$ gauge theory
supports (bulk) Majorana fermion excitations (denoted by $em$) as the bound
state of $\mathbb{Z}_2$ electric and magnetic charges.  If one condenses  these
Majorana fermions along a line segment in the lattice by lowering their excitation energy, then the Majorana zero modes will also emerge at the ends of
the chain.\cite{MajChain} Our conjecture is that the Majorana zero modes at the ends
of a Majorana chain (a line of Majorana fermion condensate) and the Majorana
zero modes associated to the extrinsic lattice dislocations are physically
equivalent. If this is true, we can create synthetic dislocations by condensing
the Majorana fermions. 

\begin{figure}[htbp]
\begin{center}
\subfigure[]{\includegraphics[width=0.128\textheight]{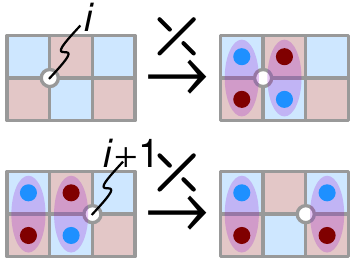}\label{fig: sigma_y action}}\qquad
\subfigure[]{\includegraphics[width=0.16\textheight]{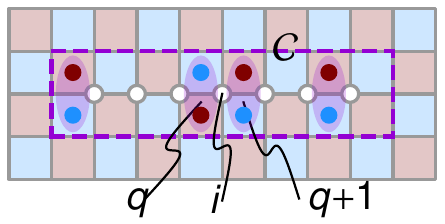}\label{fig: Majorana chain}}\\
\subfigure[]{\includegraphics[width=0.12\textheight]{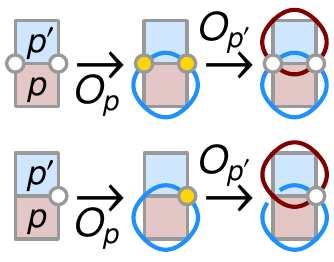}\label{fig: Z2 perturbation}}\qquad
\subfigure[]{\includegraphics[width=0.16\textheight]{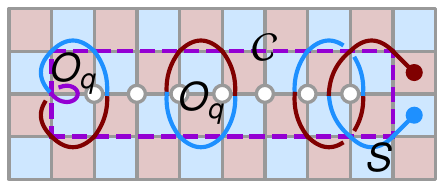}\label{fig: Z2 rings}}
\caption{(Color online.) The lattice is partitioned into even
$\protect\op{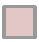}{10pt}$ and odd $\protect\op{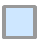}{10pt}$
plaquettes, which respectively host electric $\protect\op{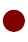}{8pt}\equiv
e$ and magnetic $\protect\op{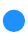}{8pt}\equiv m$ excitations. \subref{fig:
sigma_y action} The on-site action of $\sigma^y$ operator creates, annihilates
or moves fermion excitations. \subref{fig: Majorana chain} Turn on $\sigma^y$
term along a line of sites (marked by $\protect\op{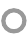}{8pt}$) to
condense the fermions into a Majorana chain. The chain (branch-cut) region
$\mathcal{C}$ is outlined by the dashed line. \subref{fig: Z2 perturbation} The
2nd order perturbation paths leading to the double-plaquette operators. Excited
sites in the intermediate states are marked by
$\protect\op{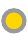}{8pt}$. \subref{fig: Z2 rings} The
double-plaquette operators on the branch-cut and around the synthetic
dislocation.}
\end{center}
\end{figure}

To verify the above conjecture, consider the $\mathbb{Z}_2$ plaquette
model on a square lattice with one qubit per site, which is described by
Hamiltonian $H_0=-\sum_p O_p$, where $O_p$ is a product of operators around the
plaquette $p$,
\begin{equation}
O_p=\op{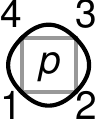}{24pt}=\sigma_1^z\sigma_2^x\sigma_3^z\sigma_4^x.
\end{equation}
Here we have adopted the graphical representation as introduced in
Ref.\,\cite{YouWen}, where each operator acting on the qubit is represented by
a string going through that site, i.e. $\sigma_i^z={\op{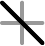}{10pt}}_i$
and $\sigma_i^x={\op{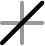}{10pt}}_i$. These operators follow the algebra
$\sigma_i^x\sigma_i^z=-\sigma_i^z\sigma_i^x$, or as
$\op{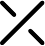}{10pt}=-\op{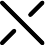}{10pt}$.  The ground state of this
model is a condensate of closed strings (generated by $O_p$ operators).
Intrinsic excitations are created by the open string operator at both of its
ends. Excitations in the even (odd) plaquettes are called electric (magnetic),
denoted by $e$ ($m$). The fermion excitation is composed of a pair of electric
and magnetic excitations in the neighboring plaquettes, which can be created
(in pair) by the action of
$\sigma_i^y=i\,{\op{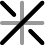}{10pt}}_i=i\sigma_i^x\sigma_i^z$ on a site, as
illustrated in \figref{fig: sigma_y action}. Further acting $\sigma_{i+1}^y$
will move one fermion to the next double plaquette
($\op{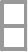}{10pt}$).  Thus the action of $\sigma_i^y$ can be
translated as $\sigma_i^y=c_{q}^\dagger c_{q+1}^\dagger+c_{q}^\dagger
c_{q+1}+h.c.$, in terms of the fermion creation $c_{q}^\dagger$  and
annihilation $c_{q}$  operators, with $q$ labeling the double plaquette, see
\figref{fig: Majorana chain}. Therefore to make a Majorana chain we only need
to add $\sigma^y$ terms to the Hamiltonian along a line of sites (in the region
$\mathcal{C}$), as $H=-\sum_p O_p+g\sum_{i\in \mathcal{C}}\sigma_i^y$. The
fermions are restricted to move along the chain, because there is no operator
like $\sigma^y$ elsewhere to take the fermions away. Each fermion excitation
carries 4 units of energy according to the $O_p$ term in the Hamiltonian $H$,
so we can write down the Hamiltonian for the fermions along the chain:
$H_\mathcal{C}=g\sum_{q\in \mathcal{C}}(c_{q}^\dagger
c_{q+1}^\dagger+c_{q}^\dagger c_{q+1}+h.c.)+4\sum_{q\in \mathcal{C}}
c_q^\dagger c_q$. For small $g$, $H_\mathcal{C}$ is in its trivial phase. For
$g$ greater than the critical value $g_c=2$, $H_\mathcal{C}$ will be driven
into the weak pairing phase,\cite{weak pairing} with Majorana zero modes at both ends of
$\mathcal{C}$. Therefore we can condense the intrinsic fermion excitations and
make a Majorana chain by applying strong ``transverse''  field $g$. This
Majorana chain will become the branch-cut line between the synthetic
dislocations.

The strong $g$ field along the branch-cut line quenches the qubit degrees of
freedom, which effectively removes those sites from the lattice, leading to the
synthetic dislocations at the ends. To justify this statement, we start from
$H=-\sum_p O_p+g\sum_{i\in \mathcal{C}}\sigma_i^y$ in the large $g$ limit. The
qubits along the branch-cut are polarized by the $g$ field to the $\sigma^y_i=-1$
state. Any plaquette operator $O_p$ that crosses the branch-cut will excite the
qubits at the crossing points to the $\sigma_i^y=+1$ state and thus taking the
system to the high energy sector (of the energy $\sim g$), as illustrated in
\figref{fig: Z2 perturbation}. Such a high energy state can be brought back to
the low energy sector immediately either by acting the same $O_p$ again or by
another plaquette operator $O_{p'}$ which shares the same crossing points.
Besides those, any other plaquette operators will leave the system in the high
energy sector, leading to higher order perturbations. So $O_p\to O_p$ and
$O_p\to O_{p'}$ are the only possible paths for perturbation up to the 2nd
order. However the former will only produce a constant shift in energy, which
is not interesting; while the later leads to the effective Hamiltonian
$H_\text{eff}=-\sum_{p\notin\mathcal{C}}O_p-1/(2g)(\sum_{q\in\mathcal{C}}O_q+2\sum_{q\in\partial\mathcal{C}}O_q)$
acting in the low energy subspace where no qubit on the branch-cut line is
excited. Here $q$ labels the double plaquettes across the branch-cut
($q\in\mathcal{C}$) or around the dislocations ($q\in\partial\mathcal{C}$). The
double-plaquette operators $O_q$ are given by
\begin{equation}
\begin{split}
O_{q\in\mathcal{C}}&=\op{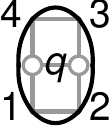}{24pt}=\sigma_1^z\sigma_2^x\sigma_3^z\sigma_4^x,\\
O_{q\in\partial\mathcal{C}}&=i\,\op{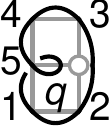}{24pt}=\sigma_1^z\sigma_2^x\sigma_3^z\sigma_4^x(-\sigma_5^y),
\end{split}
\end{equation}
whose arrangement is shown in \figref{fig: Z2 rings}. These double-plaquette
operators effectively sew up the branch-cut by coupling the qubits across, and
leaving the qubits right on the branch-cut untouched (as their degrees of
freedom has been frozen by the $g$ field). Electric and magnetic strings are
glued together in the branch-cut region by the double-plaquette operators,
which is evidenced from $[S,O_q]=0$ (take $S$ in \figref{fig: Z2 rings} for
example) such that no excitation is left on the branch cut by the action of
$S$. This explicitly demonstrate that going through the branch-cut line, $e$
charge will become $m$ and vice versa, realizing the $e$-$m$ duality, which is
the defining property of dislocation branch-cut in the $\mathbb{Z}_2$ plaquette
model.

Therefore the dislocation associated to the $e$-$m$ duality can be synthesized by condensing the fermion excitations into a Majorana chain, instead of deforming the lattice literally. The synthetic dislocations mimic all the defining physical properties of real lattice dislocations, such as implementing the $e$-$m$ duality as the intrinsic excitations winding around, and carrying the synthetic Majorana zero mode. Moreover, they are flexible and movable, as controlled by the applied field $g$, so their fusion and braiding can be discussed in a more physical sense.

\emph{Synthetic Non-Abelian Anyons}--- The above discussion can be readily generalized to the $\mathbb{Z}_N$ plaquette model,\cite{ZN1,ZN2} with one $\mathbb{Z}_N$ rotor per site, governed by the Hamiltonian $H_0=-\frac{1}{2}\sum_p O_p+h.c.$, where the single plaquette operator is defined as
\begin{equation}
O_p=\op{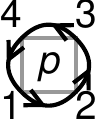}{24pt}=U_1 V_2 U_3^\dagger V_4^\dagger,
\end{equation}
with the on-site operators $U_i={\op{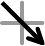}{10pt}}_i$ and $V_i={\op{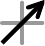}{10pt}}_i$ following the Weyl group algebra $V_iU_i=e^{i\theta_N}U_iV_i$ (with $\theta_N\equiv\frac{2\pi}{N}$), or $\op{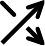}{10pt}=e^{i\theta_N}\op{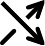}{10pt}$. Again, synthetic dislocations can be produced by simply modifying the Hamiltonian to $H=-\frac{1}{2}\sum_p O_p-\frac{1}{2}g\sum_{i\in\mathcal{C}}X_i+h.c.$, where $X_i=-e^{-i\theta_N/2}{\op{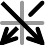}{10pt}}_i=-e^{-i\theta_N/2}U_iV_i^\dagger$. Strong enough $g$ field will quench the rotor degrees of freedom by polarization to the $X_i=+1$ state, which effectively removes the sites along the branch-cut $\mathcal{C}$. In the large $g$ limit, the low energy effective Hamiltonian follows from the 2nd order perturbation: $H_\text{eff}=-\frac{1}{2}\sum_{p\notin\mathcal{C}}O_p-1/(4g\sin^2\frac{\theta_N}{2})(\sum_{q\in\mathcal{C}}O_q+2\sum_{q\in\partial\mathcal{C}}O_q)+h.c.$, with the double-plaquette operators given by
\begin{equation}
\begin{split}
O_{q\in\mathcal{C}}&=\op{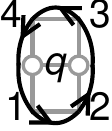}{24pt}=U_1V_2 U_3^\dagger V_4^\dagger,\\
O_{q\in\partial\mathcal{C}}&=-e^{-i\frac{\theta_N}{2}}\,\op{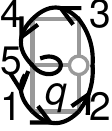}{24pt}=U_1V_2U_3^\dagger V_4^\dagger X_5.
\end{split}
\end{equation}
The double-plaquette operators glue the $\mathbb{Z}_N$ electric and magnetic
strings together, realizing the $e$-$m$ duality across the branch-cut.
Synthetic dislocations are created at both ends. It has been
shown\cite{YouWen} that each of these dislocations resembles a projective
non-Abelian anyon of quantum dimension $\sqrt{N}$.\cite{supplementary}

The point of introducing dislocations is to produce the non-Abelian anyons. In
the $\mathbb{Z}_2$ plaquette model, Majorana zero modes were produced by
condensing the fermions $em$ along a line. Now the same idea is generalized to
the $\mathbb{Z}_N$ plaquette model. From the graphical representation
$X_i=-e^{-i\theta_N/2}{\op{fig_X.pdf}{10pt}}_i$, it is clear that $X$ operator
creates, annihilates or moves the intrinsic excitations $em$ (bound states of
the $\mathbb{Z}_N$ electric and magnetic charges, which are no longer fermions
but Abelian anyons).  Strong enough $g$ field will proliferate the $em$ anyon
along the branch-cut line, dubbed as ``anyon condensation'', as the anyon
excitation energy is effectively brought to zero by the gain in kinetic energy.
The ends of the anyon chain give rise to the projective non-Abelian anyon
modes,\cite{MCheng} as a generalization of the Majorana zero modes to higher quantum
dimensions. This once again demonstrates the general principle that we can
alter the topological degeneracy and synthesize non-Abelian anyons in a
topologically-ordered Abelian system by condensing its intrinsic anyon
excitations.

\emph{Synthetic Topological Degeneracy}--- The concept of synthetic
dislocations can be further generalized by introducing the idea of
automorphism. In the previous discussion, the branch-cut line between
dislocations is like a ``magic mirror'', going through which $e$ charge becomes
$m$ charge and vice versa. Under this $e$-$m$ duality transform the statistical
properties of all intrinsic anyons are not altered. Put explicitly, in the
$\mathbb{Z}_N$ plaquette model, the self-statistic angle of $e_x m_y$ is
$e^{i\theta_N x y}$ and the mutual-statistic angle between $e_x m_y$ and
$e_{x'} m_{y'}$ is $e^{i\theta_N(xy'+x'y)}$, both angles are invariant under
$e$-$m$ duality, i.e. $x^{(\prime)}\leftrightarrow y^{(\prime)}$. Thus every
dislocation branch-cut implements a kind of automorphism (relabeling of anyons)
preserving the anyon statistics. However the $e$-$m$ duality is not the only
choice, there are also some other automorphisms, such as
$(x^{(\prime)},y^{(\prime)})\to(-x^{(\prime)},-y^{(\prime)})$, i.e. the
conjugation of both electric and magnetic charges $(e,m)\to(\bar{e},\bar{m})$.
Can they also be implemented by some kinds of branch-cuts? If so, are the
dislocations associated to any topological degeneracy? How to synthesize such
dislocations?

To answer these questions, we focus on the charge conjugate automorphism. An
$e$-string going through the charge conjugate branch-cut will become
$\bar{e}$-string with the string orientation reversed, leaving the $e_2$ anyon
(a pair of $e$) on the brach-cut line where opposite strings meet. To join the
strings, the $e_2$ anyon must be condensed along the branch-cut, such that they
are no longer excitations and can be resolved into the vacuum fluctuation, thus
the branch-cut line becomes invisible. The above design applies to the magnetic
sector as well. Therefore to synthesize charge conjugate dislocations, we only
need to proliferate the paired charges along a chain.

\begin{figure}[htbp]
\begin{center}
\subfigure[]{\includegraphics[width=0.128\textheight]{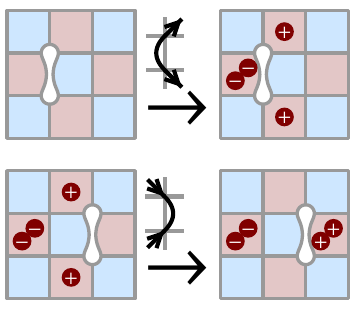}\label{fig: I action}}\qquad
\subfigure[]{\includegraphics[width=0.16\textheight]{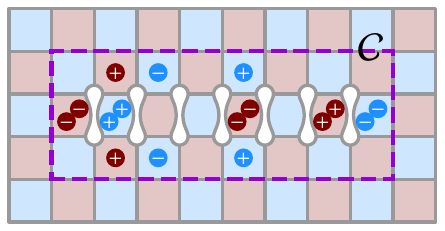}\label{fig: boson chain}}\\
\subfigure[]{\includegraphics[width=0.12\textheight]{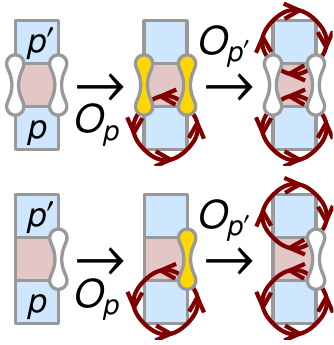}\label{fig: ZZ perturbation}}\qquad
\subfigure[]{\includegraphics[width=0.16\textheight]{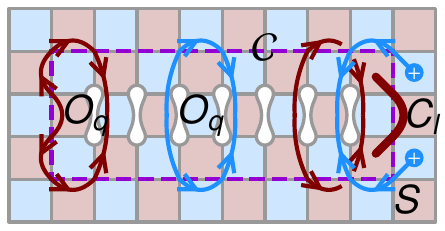}\label{fig: ZZ rings}}
\caption{(Color online.) $(\protect\op{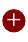}{9pt},\protect\op{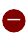}{9pt},\protect\op{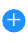}{9pt},\protect\op{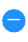}{9pt})\equiv (e,\bar{e},m,\bar{m})$ represent the $\mathbb{Z}_N$ charges. \subref{fig: I action} The on-site action of link operator $I$. \subref{fig: boson chain} Turn on $I$ term along a series of links (marked by $\protect\op{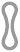}{12pt}$) to condense the anyons. \subref{fig: ZZ perturbation} The 2nd order perturbation paths leading to the triple-plaquette operators. Excited links in the intermediate states are marked by $\protect\op{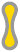}{12pt}$. \subref{fig: ZZ rings} The triple-plaquette operators on the branch-cut and around the synthetic dislocation.}
\end{center}
\end{figure}

The paired charges (either electric or magnetic) can be created, annihilated or moved by the link operator $I_l$, as illustrated in \figref{fig: I action}. $I_l$ couples two $\mathbb{Z}_N$ rotors across the link $l$:
\begin{equation}\label{eq: I}
I_l=\frac{1}{2}\left(\op{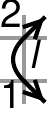}{22pt}+\op{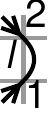}{22pt}\right)=\frac{1}{2}(U_1V_2+V_1U_2).
\end{equation}
Because $[U_1V_2,V_1U_2]=0$, all the $N^2$ eigenstates of $I_l$ are label by both eigenvalues of $U_1V_2$ and $V_1U_2$ simultaneously, which can be used to single out a unique state on which $U_1V_2=V_1U_2=1$ (and hence $I_l=+1$). Therefore strong enough $-I_l$ term can completely quench the degrees of freedom of both rotors, by polarizing the link to $I_l=+1$ state. Thus if the link operators $I$ are introduced to the the $\mathbb{Z}_N$ plaquette model along a chain of links as in \figref{fig: boson chain}, i.e. $H=-\frac{1}{2}\sum_{p}O_p-\frac{1}{2}g\sum_{l\in\mathcal{C}}I_l+h.c.$, two rows of sites (in region $\mathcal{C}$) will be effectively removed in the large $g$ limit, producing synthetic dislocations that are expected to be related to the charge conjugate automorphism according to the anyon condensation picture.

To explicitly verify the above statement, we resort to the 2nd order perturbation as shown in \figref{fig: ZZ perturbation}, and obtain the low energy effective Hamiltonian $H_\text{eff}=-\frac{1}{2}\sum_{p\notin\mathcal{C}}O_p-1/(4g\sin^2\frac{\theta_N}{2})(\sum_{q\in\mathcal{C}}O_q+2\sum_{q\in\partial\mathcal{C}}O_q)+h.c.$, with the triple-plaquette operators given by
\begin{equation}\label{eq: ZZOq}
\begin{split}
O_{q\in\mathcal{C}}&=\op{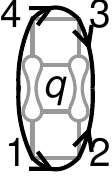}{32pt}=U_1 V_2 U_3 V_4,\\
O_{q\in\partial\mathcal{C}}&=\op{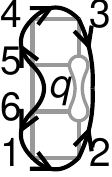}{32pt}=U_1 V_2 U_3 V_4 U_5^\dagger V_6^\dagger,
\end{split}
\end{equation}
and arranged as in \figref{fig: ZZ rings}. Because the link operator $I$ is made up of counter-oriented strings as in \eqnref{eq: I}, so any link that has been excited by a plaquette operator $O_p$ (as depicted in \figref{fig: ZZ perturbation}) can only decay (non-trivially) through the action of a counter-oriented plaquette operator $O_{p^\prime}$ across the link. This results in the convective triple-plaquette operators in \eqnref{eq: ZZOq}, which glue the opposite strings together (as exemplified by $[S,O_q]=0$ in \figref{fig: ZZ rings}), realizing the charge conjugate automorphism. Therefore dislocations (twisted defects) associated to the charge conjugate automorphism can be synthesized by a chain of link operators $I$.

For the $\mathbb{Z}_N$ plaquette model with even integer $N$, there is an additional subtlety that the effective Hamiltonian also includes the term $-\alpha \sum_{l\in\partial \mathcal{C}}C_l$ at both ends of the chain. $l\in\partial \mathcal{C}$ labels the parallel link right next to the leftmost (or rightmost) $\op{fig_linkmark0.pdf}{12pt}$-link, see \figref{fig: ZZ rings}. The operator $C_l$ is defined on such links
\begin{equation}
C_l=\left\{
\begin{array}{ll}
\op{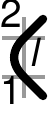}{20pt}=(-U_1 V_2)^{\frac{N}{2}} & \text{at left end},\\
\op{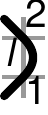}{20pt}=(-V_1 U_2)^{\frac{N}{2}} & \text{at right end}.
\end{array}\right.
\end{equation}
$C_l$ is also a local operator acting in the low energy subspace (not leading to any excitation of the energy $\sim g$), and must be presented in the effective Hamiltonian in general. It can be obtained through a $\frac{N}{2}$th order perturbation, which makes sense only for even integer $N$. However this even-odd effect should not matter in the large $N$ limit, as the coefficient $\alpha=(\frac{g}{2})^{-N/2+1}(\frac{N}{2})^{-2}$ gets exponentially weak.

Simply by counting the constrains, it is not hard to show that add each pair of such dislocations will increase the topologically protected ground state degeneracy by $(N/2)^2$ for even $N$ and by $N^2$ for odd $N$. So each dislocation has the quantum dimension $N/2$ (for even $N$) or $N$ (for odd $N$). The reasons are as follows. Applying $-g I_l$ coupling to a chain of $n$ links will quench $2n$ rotors. The rotor Hilbert space dimension is reduced by $N^{2n}$.  Meanwhile in the $\mathcal{C}$ region, the original $3(n+1)$ single-plaquette operators become $(n+1)$ triple-plaquette operators. As the ground state $|\text{grnd}\rangle$ is given by the constraints $\forall_p:O_p|\text{grnd}\rangle=|\text{grnd}\rangle$, each restricts the Hilbert space dimension by a factor of $1/N$. Under strong $g$ coupling, the number of constraints is reduced by $2(n+1)$.  The relaxation of constraints is faster than the reduction of rotor degrees of freedom, leading to an increase of the ground state degeneracy by $N^{2(n+1)}/N^{2n}=N^2$. So each dislocation is associated with the additional ground state degeneracy of $N$, hence the quantum dimension $N$. However for even integer $N$, we still have the $C_l$ operator around each dislocation. Because $C_l$ has two eigenvalues $\pm1$, it will further reduce the degeneracy by 2, resulting in the $N/2$ quantum dimension for even integer $N$. Despite of the integer quantum dimension, the charge conjugate dislocations are also non-Abelian anyons.\cite{supplementary}

\emph{Conclusion}--- In this work, we proposed the microscopic construction of flexible and movable synthetic dislocations by simply modifying the Hamiltonian along a line of sites or links. Two kinds of synthetic dislocations in the $\mathbb{Z}_N$ plaquette model were explicitly constructed: one associated to the $e$-$m$ duality, and the other associated to the charge conjugation. Both are automorphisms among the intrinsic anyons related by statistical symmetry. A systematic way to design synthetic dislocations is to condense the intrinsic anyon excitations that is left over by the automorphism along the branch-cut line. The $e$-$m$ duality (charge conjugate) dislocations can be synthesized by condensing a chain of $em$ anyon (paired charges). In principle, even more general dislocations can be synthesized by arbitrary anyon condensation.
The synthetic dislocations are associated to additional topological degeneracy, and rendered as projective non-Abelian anyons. Therefore anyon condensation provides us a controllable and systematic way to generate the topologically protected ground state degeneracy in a topologically ordered system.

\begin{acknowledgments}
We thank Maissam Barkeshli and Xiao-Liang Qi for the discussion on the application of loop algebra approach developed in Ref.\,\cite{BarkeshliJianQi}, and Liang Kong for helpful discussions. This work is supported by NSF Grant No. DMR-1005541 and NSFC 11074140.
\end{acknowledgments}

\onecolumngrid
\section{Supplementary Material}
\begin{figure}[htbp]
\begin{center}
\includegraphics[width=0.18\textheight]{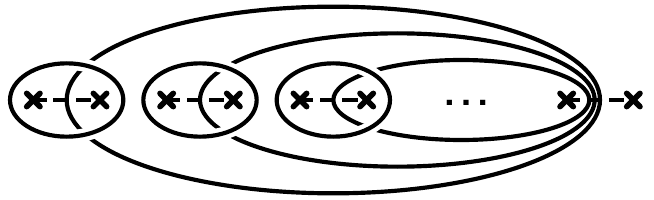}
\caption{Non-contractable Wilson loops are chosen according to the homology basis, following Ref.\,\cite{BarkeshliWen}.}
\label{fig: Wilson loops}
\end{center}
\end{figure}

\twocolumngrid
In the supplementary material, we discuss the anyonic properties of the synthetic dislocations constructed in this work, including their quantum dimensions and braiding rules. Because these properties are not sensitive to the microscopic details, they can be studied in the continuous space using the loop algebra approach developed by Barkeshli, Jian and Qi \cite{BarkeshliJianQi}, which can be considered as a long-wave-length effective theory of synthetic dislocations. The essential idea of loop algebra is to use non-contractable Wilson loops (closed strings of operators) to keep tract of the dislocation configurations in the system and specify the ground state Hilbert space, such that the effect of adding, removing or braiding the dislocations can all be handled by the algebraic relations among these Wilson loop operators. If we have several pairs of synthetic dislocations, we may arrange them along a line and choose the non-contractable loops like those in \figref{fig: Wilson loops}. Adding each pair of dislocations into the system will introduce two additional loops: one going around the pair, and the other connects to the last pair of dislocations. All the Wilson loop operators   commute with the Hamiltonian by definition, so the ground state Hilbert space is simply a representation space (presumably irreducible) of the loop algebra. The new loops introduced by new dislocations will enlarge the representation space, and hence increase the ground state degeneracy. As has been pointed out in Ref.\,\cite{BarkeshliJianQi}, braiding two dislocations leads to the deformation of the Wilson loops, so the ground states specified by these loops are also altered, resulting in a non-Abelian Berry phase that gives rise to the projective non-Abelian statistics of dislocations. All these general principles will be demonstrated with definite examples from $\mathbb{Z}_N$ plaquette model in the following.

There are two strings in the $\mathbb{Z}_N$ plaquette model: electric $\op{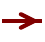}{10pt}$ and magnetic $\op{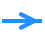}{10pt}$, following the local algebra $\op{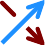}{10pt}=e^{i\theta_N}\op{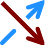}{10pt}$ (with $\theta_N\equiv\frac{2\pi}{N}$), $\op{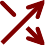}{10pt}=\op{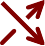}{10pt}$ and $\op{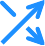}{10pt}=\op{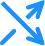}{10pt}$. Dislocations are characterized by their stabilizers (the plaquette operators at the end of the branch-cut line). According the microscopic construction, the stabilizer of $e$-$m$ duality dislocation reads $O_q=-e^{-i\frac{\theta_N}{2}}\op{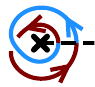}{18pt}$, and the stabilizer of charge conjugate dislocation reads $O_q=\op{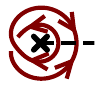}{18pt}$ or $O_q=\op{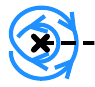}{18pt}$. The ground state is given by $O_q=+1$, which means in the ground state manifold, we have
\begin{equation}\label{eq: stabilizer}
\begin{array}{ll}
\op{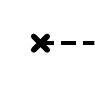}{18pt}=-e^{-i\frac{\theta_N}{2}}\op{appfig_OOem.pdf}{18pt}& \text{for $e$-$m$ duality,}\\
\op{appfig_dis.pdf}{18pt}=\op{appfig_OOee.pdf}{18pt}=\op{appfig_OOmm.pdf}{18pt} & \text{for charge conjugate.}
\end{array}
\end{equation}
In the following, we will focus on the case of having 4 dislocations (or 2 pairs) resting on a sphere. Denote the braiding of dislocations 1 and 2 as $\sigma_{12}$, and the braiding of 2 and 3 as $\sigma_{23}$. On the sphere, braiding 3 and 4 is equivalent to braiding 1 and 2, and thus will not be discussed.

\begin{figure}[htbp]
\begin{center}
\subfigure[]{\includegraphics[width=0.12\textheight]{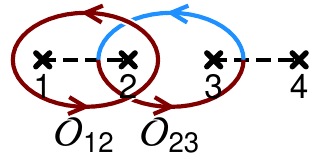}\label{fig: em loops}}\qquad
\subfigure[]{\includegraphics[width=0.12\textheight]{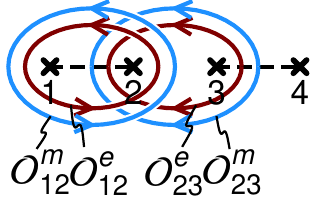}\label{fig: cc loops}}
\caption{(Color online.) Non-contractable Wilson loops for 4 dislocations on a sphere in the case of \subref{fig: em loops} $e$-$m$ duality dislocations, \subref{fig: cc loops} charge conjugate dislocations.}
\end{center}
\end{figure}

We start from the $e$-$m$ duality dislocations, as shown in \figref{fig: em loops}. $\mathcal{O}_{12}$ and $\mathcal{O}_{23}$ are the Wilson loops, following the algebra $\mathcal{O}_{23}\mathcal{O}_{12}=e^{i\theta_N}\mathcal{O}_{12}\mathcal{O}_{23}$, whose irreducible representation space is of $N$-dimension, with the basis specified by ($n=0,1,\cdots,N-1$)
\begin{equation}\label{eq: emOO}
\mathcal{O}_{12}\ket{n}=e^{in\theta_N}\ket{n}, \mathcal{O}_{23}\ket{n}=\ket{n-1}.
\end{equation}
Because $[\mathcal{O}_{12},H]=[\mathcal{O}_{23},H]=0$, the Hamiltonian operator can be represented as $H=0$ in the representation space, thus the states $\ket{n}$ represent the degenerated ground states of the Hamiltonian. Therefore adding each pair of $e$-$m$ duality dislocations into the system will increase the ground degeneracy by a factor of $N$, hence each dislocation is of quantum dimension $\sqrt{N}$.\cite{YouWen}

Braiding the dislocations will deform the Wilson loops. Under $\sigma_{12}$, $\mathcal{O}_{12}$ is not affected, however $\mathcal{O}_{23}$ undergoes the following deformation
\begin{equation}
\begin{split}
\mathcal{O}_{23}=&\graph{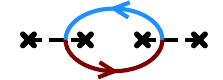}\\
\overset{\sigma_{12}}{\longrightarrow}&\graph{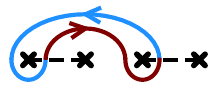}\\
=&-e^{-i\frac{\theta_N}{2}}\graph{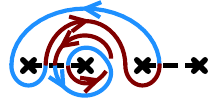}\\
=&-e^{-i\frac{\theta_N}{2}}\graph{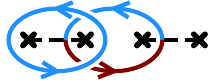}\\
=&-e^{-i\frac{\theta_N}{2}}\graph{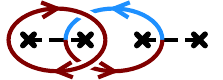}\\
=&-e^{-i\frac{\theta_N}{2}}\mathcal{O}_{12}^\dagger\mathcal{O}_{23},
\end{split}
\end{equation}
where we have used \eqnref{eq: stabilizer} to create a stabilizer ring around the dislocation 2. Under $\sigma_{23}$, $\mathcal{O}_{23}$ is not affected, however $\mathcal{O}_{12}$ undergoes the following deformation
\begin{equation}
\begin{split}
\mathcal{O}_{12}=&\graph{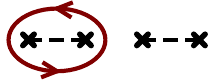}\\
\overset{\sigma_{23}}{\longrightarrow}&\graph{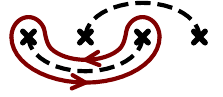}\\
=&\graph{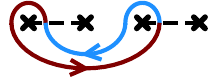}\\
=&-e^{-i\frac{\theta_N}{2}}\graph{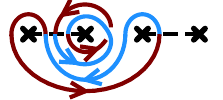}\\
=&-e^{-i\frac{\theta_N}{2}}\graph{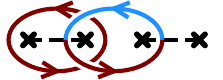}\\
=&-e^{-i\frac{\theta_N}{2}}\mathcal{O}_{23}\mathcal{O}_{12},
\end{split}
\end{equation}
where we have rearranged the branch-cut lines between dislocations after the braiding, and used \eqnref{eq: stabilizer} to insert the stabilizer. In summary, under the braiding of dislocations, the Wilson loops transform as
\begin{equation}
\begin{split}
&\left\{\begin{array}{l}
\mathcal{O}_{12}\overset{\sigma_{12}}{\longrightarrow}\mathcal{O}_{12},\\
\mathcal{O}_{23}\overset{\sigma_{12}}{\longrightarrow}-e^{-i\frac{\theta_N}{2}}\mathcal{O}_{12}^\dagger\mathcal{O}_{23},
\end{array}\right.\\
&\left\{\begin{array}{l}
\mathcal{O}_{12}\overset{\sigma_{23}}{\longrightarrow}-e^{-i\frac{\theta_N}{2}}\mathcal{O}_{23}\mathcal{O}_{12},\\
\mathcal{O}_{23}\overset{\sigma_{23}}{\longrightarrow}\mathcal{O}_{23}.
\end{array}\right.
\end{split}
\end{equation}

Now we seek the unitary transform $T_{12}$ and $T_{23}$ in the ground state Hilbert space to represent $\sigma_{12}$ and $\sigma_{23}$ respectively, such that
\begin{equation}\label{eq: emTOT}
\begin{split}
&\left\{\begin{array}{l}
T_{12}^\dagger\mathcal{O}_{12}T_{12}=\mathcal{O}_{12},\\
T_{12}^\dagger\mathcal{O}_{23}T_{12}=-e^{-i\frac{\theta_N}{2}}\mathcal{O}_{12}^\dagger\mathcal{O}_{23},
\end{array}\right.\\
&\left\{\begin{array}{l}
T_{23}^\dagger\mathcal{O}_{12}T_{23}=-e^{-i\frac{\theta_N}{2}}\mathcal{O}_{23}\mathcal{O}_{12},\\
T_{23}^\dagger\mathcal{O}_{23}T_{23}=\mathcal{O}_{23}.
\end{array}\right.
\end{split}
\end{equation}
Given the representation of $\mathcal{O}_{12}$ and $\mathcal{O}_{23}$ in \eqnref{eq: emOO}, it can be verified that the following is a solution of \eqnref{eq: emTOT}, up to U(1) phase factors for both $T_{12}$ and $T_{23}$,
\begin{equation}
\begin{split}
T_{12}&=\sum_{n}e^{i\frac{\theta_N}{2}n(N-n)}\ket{n}\bra{n},\\
T_{23}&=\frac{e^{-i\pi/4}}{\sqrt{N}}\sum_{n,n'}e^{i\frac{\theta_N}{2}(n-n'-N/2)^2}\ket{n}\bra{n'}.
\end{split}
\end{equation}
One can check that $T_{12}$ and $T_{23}$ satisfy the braid group definition relation: $T_{12}T_{23}T_{12}=T_{23}T_{12}T_{23}$. As braiding the dislocations deforms the Wilson loops, the ground states defined by the deformed Wilson loops after braiding is related to the original ground states by the unitary transform $T_{ab}$, so $T_{ab}$ represents the non-Abelian Berry phase accumulated during the braiding. Because there is a U(1) phase ambiguity as \eqnref{eq: emTOT} is symmetric under the transform $T_{ab}\to e^{i\theta_{ab}}T_{ab}$, hence the braid group definition relation only holds up to a U(1) phase $e^{i\phi}$, i.e. $T_{12}T_{23}T_{12}=e^{i\phi} T_{23}T_{12}T_{23}$, so the dislocations are projective non-Abelian anyons.

We continue to consider the charge conjugate dislocations, as shown in \figref{fig: cc loops}. Adding each pair of such dislocations will introduce four non-contractible Wilson loops: $\mathcal{O}_{12}^e$, $\mathcal{O}_{23}^e$ of electric type, and $\mathcal{O}_{12}^m$, $\mathcal{O}_{23}^m$ of magnetic type. They can be separated into two independent sets of algebra $\mathcal{O}_{23}^m\mathcal{O}_{12}^e=e^{2i\theta_N}\mathcal{O}_{12}^e\mathcal{O}_{23}^m$, $\mathcal{O}_{23}^e\mathcal{O}_{12}^m=e^{2i\theta_N}\mathcal{O}_{12}^m\mathcal{O}_{23}^e$ and $[\mathcal{O}_{12}^e,\mathcal{O}_{23}^e]=[\mathcal{O}_{12}^m,\mathcal{O}_{23}^m]=[\mathcal{O}_{12}^e,\mathcal{O}_{12}^m]=[\mathcal{O}_{23}^e,\mathcal{O}_{23}^m]=0$. If we use the common eigenstates of $\mathcal{O}_{12}^e$ and $\mathcal{O}_{12}^m$ as the basis of the ground state Hilbert space
\begin{equation}\label{eq: ccOO1}
\begin{split}
\mathcal{O}_{12}^e\ket{n_e,n_m}&=e^{in_e\theta_N}\ket{n_e,n_m},\\
\mathcal{O}_{12}^m\ket{n_e,n_m}&=e^{in_m\theta_N}\ket{n_e,n_m},
\end{split}
\end{equation}
where $n_e$ and $n_m$ are two integers introduced to label the ground states, then $\mathcal{O}_{23}^{m(e)}$ simply lowers the index $n_{e(m)}$ by 2 (modulo $N$) according to the algebraic relations mentioned above
\begin{equation}\label{eq: ccOO2}
\begin{split}
\mathcal{O}_{23}^m\ket{n_e,n_m}&=\ket{n_e-2,n_m},\\
\mathcal{O}_{23}^e\ket{n_e,n_m}&=\ket{n_e,n_m-2}.
\end{split}
\end{equation}
At this point it becomes obvious why the parity of $N$ is important. If $N$ is even, starting from a $n_e=0$ state, the operator $\mathcal{O}_{23}^m$ can only take it to the states labeled by even integer $n_e$ between 0 and $N-2$, i.e. $n_e=0,2,4,\cdots,(N-2)$. However if $N$ is odd, then all the states with integer $n_e$ between 0 and $N-1$ can be achieved by acting $\mathcal{O}_{23}^m$, i.e.   $n_e=0,2,4,\cdots,(N-1),1,3,5,\cdots,(N-2)$. So for even integer $N$, both $n_e$ and $n_m$ only take $N/2$ even integer values, and the ground state Hilbert space associated to a pair of dislocations is of $(N/2)^2$-dimension; while for odd integer $N$, both $n_e$ and $n_m$ can take $N$ integer values, thus the associated ground state Hilbert space is of $N^2$-dimension. For later convenience, we introduce the notation $\db{N}$
\begin{equation}
\db{N}=\left\{\begin{array}{ll}N/2& \text{if $N$ even},\\N&\text{if $N$ odd}.\end{array}\right.
\end{equation}
Then the quantum dimension associated to each charge conjugate dislocation is simply $\db{N}$.

Consider braiding the charge conjugate dislocations, we analyze the deformation of the $e$-loops first. The analysis of the $m$-loops will be essentially the same. Under $\sigma_{12}$, $\mathcal{O}_{12}^e$ is not affected, however $\mathcal{O}_{23}^e$ undergoes the following deformation
\begin{equation}
\begin{split}
\mathcal{O}_{23}^e=&\graph{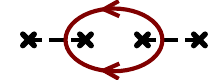}\\
\overset{\sigma_{12}}{\longrightarrow}&\graph{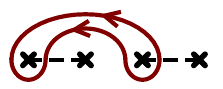}\\
=&\graph{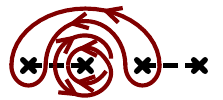}\\
=&\graph{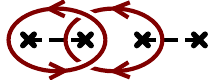}\\
=&\mathcal{O}_{12}^e\mathcal{O}_{23}^e,
\end{split}
\end{equation}
where the stabilizer is inserted according to \eqnref{eq: stabilizer}. Under $\sigma_{23}$, $\mathcal{O}_{23}^e$ is not affected, however $\mathcal{O}_{12}$ undergoes the following deformation,
\begin{equation}
\begin{split}
\mathcal{O}_{12}^e=&\graph{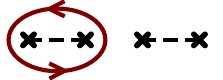}\\
\overset{\sigma_{23}}{\longrightarrow}&\graph{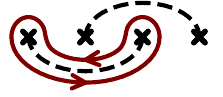}\\
=&\graph{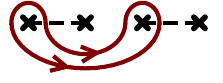}\\
=&\graph{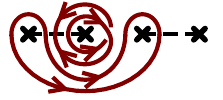}\\
=&\graph{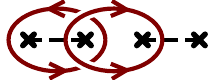}\\
=&\mathcal{O}_{23}^{e\dagger}\mathcal{O}_{12}^e,
\end{split}
\end{equation}
where we have rearranged the branch-cut lines between dislocations after the braiding, and used \eqnref{eq: stabilizer} to insert the stabilizer. The $m$-loops deforms in the same way as $e$-loops, and will not be elaborated again. In summary, under the braiding of dislocations, the Wilson loops transform as
\begin{equation}
\begin{split}
&\left\{\begin{array}{l}
\mathcal{O}_{12}^{e(m)}\overset{\sigma_{12}}{\longrightarrow}\mathcal{O}_{12}^{e(m)},\\
\mathcal{O}_{23}^{e(m)}\overset{\sigma_{12}}{\longrightarrow}\mathcal{O}_{12}^{e(m)}\mathcal{O}_{23}^{e(m)},
\end{array}\right.\\
&\left\{\begin{array}{l}
\mathcal{O}_{12}^{e(m)}\overset{\sigma_{23}}{\longrightarrow}\mathcal{O}_{23}^{e(m)\dagger}\mathcal{O}_{12}^{e(m)},\\
\mathcal{O}_{23}^{e(m)}\overset{\sigma_{23}}{\longrightarrow}\mathcal{O}_{23}^{e(m)}.
\end{array}\right.
\end{split}
\end{equation}

Again we seek the unitary transform $T_{12}$ and $T_{23}$ in the ground state Hilbert space to represent $\sigma_{12}$ and $\sigma_{23}$ respectively, such that
\begin{equation}\label{eq: ccTOT}
\begin{split}
&\left\{\begin{array}{l}
T_{12}^\dagger\mathcal{O}_{12}^{e(m)}T_{12}=\mathcal{O}_{12}^{e(m)},\\
T_{12}^\dagger\mathcal{O}_{23}^{e(m)}T_{12}=\mathcal{O}_{12}^{e(m)}\mathcal{O}_{23}^{e(m)},
\end{array}\right.\\
&\left\{\begin{array}{l}
T_{23}^\dagger\mathcal{O}_{12}^{e(m)}T_{23}=\mathcal{O}_{23}^{e(m)\dagger}\mathcal{O}_{12}^{e(m)},\\
T_{23}^\dagger\mathcal{O}_{23}^{e(m)}T_{23}=\mathcal{O}_{23}^{e(m)}.
\end{array}\right.
\end{split}
\end{equation}
Note that the above equations should hold for both $\mathcal{O}_{ab}^e$ and $\mathcal{O}_{ab}^m$ simultaneously. Given the representation of $O_{ab}^{e(m)}$ in Eq.\,(\ref{eq: ccOO1},\ref{eq: ccOO2}), it is easy to show that the following is a solution of \eqnref{eq: ccTOT}, up to U(1) phase factors,
\begin{equation}\label{eq: ccT1T2}
\begin{split}
T_{12}&=\sum_{n_e,n_m}e^{in_en_m\theta_N/2}\ket{n_e,n_m}\bra{n_e,n_m},\\
T_{23}&=\frac{1}{\db{N}}\sum_{n_e,n_m,n_e^\prime,n_m^\prime}e^{-i(n_e-n_e^\prime)(n_m-n_m^\prime)\theta_N/2}\\
&\hspace{90pt}\ket{n_e,n_m}\bra{n_e^\prime,n_m^\prime}.
\end{split}
\end{equation}
To avoid complicated piecewise formulation of the even-odd effect, $n_e,n_m,n_e^\prime,n_m^\prime$ should take their values from $0,2,4,\cdots,2(\db{N}-1)$. For example, if $N=5$, then $n_{e(m)}^{(\prime)}=0,2,4,6,8$; and if $N=6$, then $n_{e(m)}^{(\prime)}=0,2,4$. The state $\ket{n_e,n_m}$ should be understood as $\ket{n_e\mod N, n_m\mod N}$ (under modulo $N$). Then \eqnref{eq: ccT1T2} is applicable to both even and odd $N$. One can check that $T_{12}$ and $T_{23}$ satisfy the braid group definition relation: $T_{12}T_{23}T_{12} = T_{23}T_{12}T_{23}$. Thus $T_{ab}$ is the non-Abelian Berry phase (upto a U(1) phase) accumulated during the braiding of dislocations $a$ and $b$. Therefore the charge conjugate dislocations are also projective non-Abelian anyons.

\end{document}